\documentclass[twocolumn,showpacs,preprintnumbers,amsmath,amssymb]{revtex4}

\usepackage{graphicx}
\usepackage{dcolumn}
\usepackage{bm}

\begin{document}

\title{First-passage times for random walks in bounded domains}

\author{S. Condamin}
\author{O. B\'enichou}
\author{M. Moreau}
\affiliation{Laboratoire de Physique Th\'eorique de la Mati\`ere Condens\'ee 
(UMR 7600), case courrier 121, Universit\'e Paris 6, 4 Place Jussieu, 75255
Paris Cedex  }

\date{\today}

\begin{abstract}

We present a novel computational method of first-passage times between a starting site and a target site of regular bounded lattices. 
We derive  accurate expressions for all the moments of this first-passage time, validated by  numerical simulations.  Their range of validity is discussed. We also consider the case of a starting site and two targets.  In addition, we present the extension to continuous Brownian motion. These results are of great relevance to any system involving diffusion in confined media.
\end{abstract}


\maketitle


How long does it take for a drunkard to go from a given bar to another one? 
This time is known in the random walk literature as 
a first passage time (FPT), and it has generated  a considerable amount of 
work for many years \cite{VanKampen,Redner}. The importance of  
FPT relies on 
the fact that many physical properties, including fluorescence quenching 
\cite{Rice}, 
neuron dynamics \cite{Tuckwell} or resonant activation \cite{Doering} to name 
a few, are controlled 
by first passage events.  Unfortunately, explicit determinations of FPT are most of the time limited to  very artificial geometries, such as 1D and spherically symmetric problems \cite{Redner}.

The determination of FPT for random walks in realistic geometries is not just a theoretical challenge in its own right. 
 It is actually a very general issue involved  as soon as molecules diffuse in confined media  as, for example, biomolecules diffusing in the cell and  undergoing a series of transformations   at precise regions of the cell. An estimation of the time needed to go from one point to another is then an essential step in the understanding of the kinetics of the whole process.



Very recently, two important advances in the calculation of FPT have been performed. First, in 
the case of discrete random walks, an expression for the mean first passage 
time (MFPT) between two nodes of a general network has been found \cite{Noh}.
So far, however, no quantitative estimation of the MFPT has been derived  from 
this formula.
Second,  the leading behavior of  MFPT  of a continuous Brownian motion at a small absorbing window of a general reflecting bounded domain  has been obtained \cite{Holcman05,bere}.  In the case when this window is 
a small sphere within the domain, the  behavior of MFPT has 
recently been derived \cite{Pinsky}. This result is rigorous, but does not give access to the dependence of the MFPT with the starting site.

In this letter, we present a new computational method that allows us  to 
quantitatively  extend all these results in three directions: (i) we obtain 
an accurate explicit formula for the MFPT, (ii) we   
also examine all the moments of the FPT, and (iii) we consider the 
case with two  targets. The method is presented in detail in the case of 
discrete random walks on regular lattices, and then the extension to continuous
 Brownian motion is outlined.     

We first consider a random walker on a bounded lattice, and we address
the  question of determining the mean time needed to reach one point
of the lattice  (target site $T$) from another one (starting site $S$). The
boundaries are assumed to  be reflecting. The starting point of
the method  is a result known  in the mathematical literature as
Kac's formula \cite{Aldous}.  Indeed, after our previous work about
first return times (FRTs) for random walks  \cite{condamin}, we found out that
Kac's formula allows one to extend our results  to general finite
graphs. Kac's result concerns   irreductible graphs (ie 
 from any point one can go to any other point),  which   admit a
stationary probability $\pi({\bf r})$ to be at site ${\bf r}$. Let us  consider random walks
starting  from a random point of a subset $\Sigma$ of sites of the lattice, with a probability proportional to the
stationary probability. Then, the mean
FRT of the random walk, i.e. the mean number of steps needed to return to any point
of $\Sigma$,   is, according to Kac's formula,   $1/\pi(\Sigma)$, where $\pi(\Sigma) =
\sum_{{\bf r} \in \Sigma} \pi({\bf r})$. 

\begin{figure}[t]
\centering\includegraphics[width = .55\linewidth]{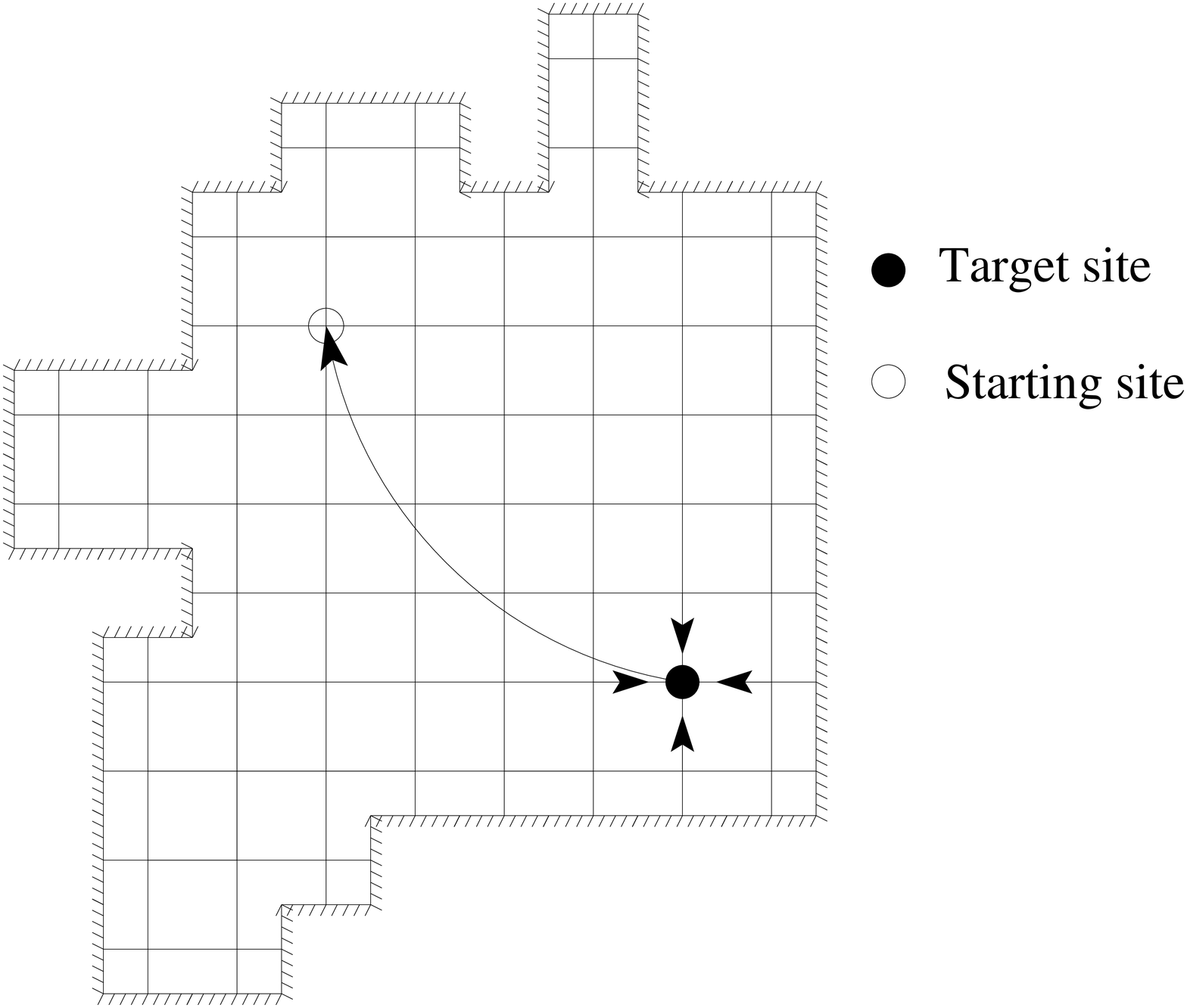}
\caption{Modifications of the original lattice: arrows denote one-way links.}
\label{astuce}
\end{figure}

This formula gives FRTs and not FPTs. However, we can use it to derive
the MFPT $\langle \mathbf{T}\rangle $ by slightly  modifying the
original lattice  (see fig.\ref{astuce}):  we suppress all the
original links starting from the target site $T$, and add a new
one-way link from $T$ to the starting point $S$. In this modified
lattice, 
the FRT to $T$ is just the FPT from $S$ to $T$ in the
former lattice, plus one.
In what follows, for sake of simplicity, we  only consider regular 2D
or 3D lattices, although the argument may be easily extended to any
kind of graph.  Let ${\bf r}_T$ be the position of the target site,
and ${\bf r}_S$ be the  position of the starting site.  Denoting $
\pi({\bf r}_T) = j$, Kac's formula gives  $ \langle \mathbf{T}\rangle
= 1/j - 1$.  Thus, all we need to  know is thus the stationary
probability for the modified graph.  It  satisfies the following
equation:
\begin{equation}
\label{stat}
\pi({\bf r}) = \frac1\sigma \sum_{\langle{\bf r}',{\bf
r}\rangle}\pi({\bf r}') +  j\delta_{{\bf r}{\bf r}_S} -
\sum_{\langle{\bf r}',{\bf r}_T\rangle} \frac{j}{\sigma} \delta_{{\bf
r}{\bf r}'}
\end{equation}
where $\langle{\bf r},{\bf r}'\rangle$ means that these two sites
are neighbors and   $\sigma$ is the number of nearest neighbors of a
site (by convention, the sites on the boundaries are their own
neighbors). The last two terms of the rhs of (\ref{stat}) are due to the modifications of the lattice.   To solve this equation, we define the auxiliary function
$\pi'$,  equal to $\pi$ for ${\bf r} \neq {\bf r}_T$, with $\pi'({\bf
r}_T) = 0$.  It satisfies:
\begin{equation}
\pi'({\bf r}) = \frac1\sigma \sum_{\langle{\bf r}',{\bf
 r}\rangle}\pi'({\bf r}') +  j\delta_{{\bf r}{\bf r}_S} -
 j\delta_{{\bf r}{\bf r}_T},
\label{eqpip}
\end{equation}
so that $\pi'$ has the following expression:
\begin{equation}
\pi'({\bf r}) = \frac{1-j}N + j H({\bf r}|{\bf r}_S) - j H({\bf
r}|{\bf r}_T),
\label{valpip}
\end{equation}
where $N$ is the total number of sites, and $H$ is the discrete
pseudo-Green function \cite{Barton}, which is symmetrical  in its
arguments and satisfies:
\begin{equation}
H({\bf r}|{\bf r}') = \frac{1}{\sigma}\sum_{\langle{\bf r}'',{\bf
r}\rangle}  H({\bf r}''|{\bf r}')+ \delta_{{\bf rr}'} - \frac1N
\label{pseudogreen}
\end{equation}
Indeed the solution (\ref{valpip}) satisfies equation (\ref{eqpip}),
and  ensures that $\pi$ is a probability function (of sum unity).  The
condition $\pi'({\bf r}_T)=0$ allows us to compute $j$ and to deduce
the following exact expression:
\begin{equation}
\langle \mathbf{T}\rangle = N [ H({\bf r}_T|{\bf r}_T) - H({\bf
r}_T|{\bf r}_S)  ]
\end{equation}
This formula is equivalent to the one found in \cite{Noh}, but is
expressed in  terms of pseudo-Green functions. One advantage of our
method is that it  may  be easily extended to more complex situations,
as we will show.   Another advantage is that, although the
pseudo-Green function $H$ is not known  in general, it is well suited
to approximations.  The simplest one is to approximate the
pseudo-Green function by its  infinite-space limit, the "usual" Green
function: $H({\bf r}|{\bf r}')\simeq  G({\bf r}-{\bf r}')$, which
satisfies:
\begin{equation}
G({\bf r}) = \frac{1}{\sigma}\sum_{\langle{\bf r}',{\bf r}\rangle}
G({\bf r}')+ \delta_{0{\bf r}}.
\label{green}
\end{equation} 
The value of $G(0)$ and the asymptotic behaviour of $G$ are well-known
\cite{Hughes}. For instance, for the 3D cubic lattice, we have:  $G(0)
= 1.516386$  and $G({\bf r}) \simeq 3/(2\pi r)$ for $r$ large.
For the 2D square lattice, we have  $G(0)-G({\bf r}) \simeq
2/\pi \ln(r) + 3/\pi \ln2 +  2\gamma/\pi$, where
$\gamma$ is the Euler gamma constant.   These estimations of $G$ are
used for all the practical applications in the following.  This
infinite-space approximation may be improved by two kinds of
corrections.  First,  the constant term $1/N$ in equation
(\ref{pseudogreen}) may be taken into account:
\begin{equation}
H({\bf r}|{\bf r}')\simeq G({\bf r}-{\bf r}') + \frac1N ({\bf r}-{\bf
r}')^2
\label{eqh}
\end{equation}
In the 3D case, this ``uniform correction'' is always weak: its order
of  magnitude is at most $N^{-1/3}$. However, in the 2D case, it is
negligible only if $N\ln({\bf r}_S-{\bf r}_T) \gg ({\bf r}_S - {\bf
r}_T)^2$.

A second correction that may be taken into account is the influence of
nearby boundaries. For flat boundaries, it can be  computed
explicitly.  Denoting by $s({\bf r})$  the symmetric point of site
${\bf r}$  with respect to the closest flat boundary, $H$ becomes:
\begin{equation}
H({\bf r}|{\bf r}')\simeq G({\bf r}-{\bf r}') + G({\bf r}-s({\bf r}'))
\label{eqh2}
\end{equation}
If the boundary is not flat, this expression only gives the order of
magnitude of the  expected correction. These two alternative
corrections correspond to two  different ways to treat the effect of
boundaries:  (\ref{eqh}) is a mean-field type correction, whereas
(\ref{eqh2}) is a local correction.  One should use either (\ref{eqh})
or (\ref{eqh2}) mainly according to the  position of the target. A
rule of thumb, used in the following,  is that  as soon as one of the
two corrections is negligible, the other one leads to  good results.
Indeed, the correction (\ref{eqh}) is useful for a target far from
any boundary, whereas the correction (\ref{eqh2}) is  more appropriate
if the target is close to a flat boundary. As for the limitations of
these approximations, they are not to be used in two cases: (i) if
neither (\ref{eqh}) nor (\ref{eqh2}) are negligible; 
(ii) if the target is close to an irregular boundary.

\begin{figure}[t]
\centering\includegraphics[width = .7\linewidth,clip]{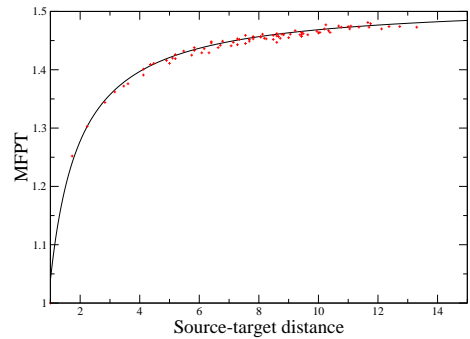}
\caption{3D - Influence of the distance between the source and 
the
target. Simulations (red crosses) vs. theory (plain line). The domain
is a cube of side 31, the target being in the middle of it. The source
takes all the positions in a cube of side 15 centered on the target.}
\label{distst}
\end{figure}
\begin{figure}[t]
\centering\includegraphics[width = .7\linewidth,clip]{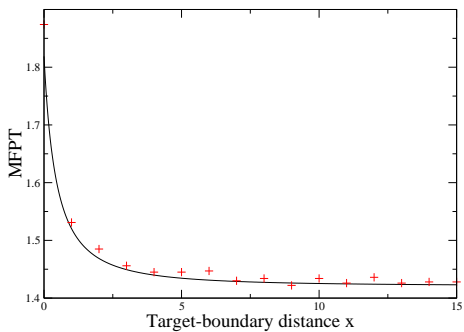}
\caption{3D - Influence of the distance between the target and 
a flat boundary. Simulations (red crosses) vs. theory (plain line). The
domain is a cube of side 41 whose center is at (0,0,0); the source  is
at (0,0,x-15) and the target is at (0,0,x-20).}
\label{disttw}
\end{figure}

We have compared the theoretical predictions with numerical
simulations.  We first checked (fig. \ref{distst}) the behavior  of
the MFPT when  the source-target distance varies (the FPT  is averaged
over 100 000 random walks).  We also studied the influence of the
distance between the target and a  boundary (fig.\ref{disttw}), using
the correction (\ref{eqh2}).
\begin{figure}[t]%
\centering\includegraphics[width = 0.7\linewidth,clip]{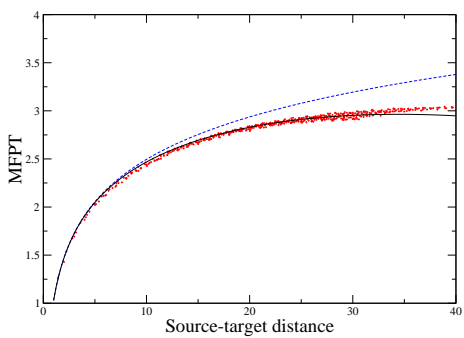}
\caption{2D - Influence of the distance between the source  
and the target. Simulations (red crosses) vs. theory: blue dotted line,
without the uniform correction; black plain line 
(in the middle of the set of points) 
, with this correction.  The domain is a square of side 61, the target being in
the middle of it, and the source takes all possible positions.}
\label{distst2D}
\end{figure}%
Finally, we checked that our approximation was also correct for the 2D
case (fig.\ref{distst2D}).  Since in this case the uniform correction
(\ref{eqh}) is not negligible, we took it into account.

In all the cases studied, the numerical simulations validate our
approximations.  Thus, our method provides an efficient way for
estimating the MFPT,  which up to now was only known formally and for
a few specific cases.

\begin{figure}[t]
\centering\includegraphics[width = .7\linewidth,clip]{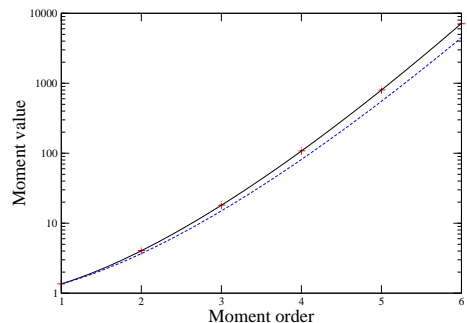}
\caption{3D - Higher-order moments: Theory (black curve) vs. 
simulation (red 
crosses). The blue dotted curve is the moments of the exponential 
distribution whose average time is the MFPT.
The $n$-th moment is normalized
by $N^n$; the domain is a cube of side 51 centered on the target at (0,0,0) and the
source at (2,2,1).}
\label{moments}
\end{figure}

Furthermore, it is possible  to compute the higher-order moments  of
the FPT, using an extension of Kac's formula, which gives a relation
between the Laplace transform of the FRT to a subset $\Sigma$, averaged 
on $\Sigma$, and the Laplace transform of the FPT to this same subset, 
averaged on the complementary subset $\bar{\Sigma}$.  
\begin{equation}
\pi(\Sigma)\left( \left\langle e^{-s\mathbf{T}}\right\rangle_\Sigma -
e^{-s} \right) =  (1 -\pi(\Sigma))\left(e^{-s} - 1\right) \left\langle
e^{-s\mathbf{T}} \right\rangle_{\bar{\Sigma}} \nonumber
\end{equation}
Both averages are weighted by the stationary probability $\pi$.
For any starting point $M$ different from $T$, the behaviour 
of the random walk from $M$ to $T$ is exactly the same on the original and 
modified lattices until it reaches $T$. Thus, the FPT from $M$ to $T$ is 
the same on both lattices. This remark allows one, by applying the above 
formula
to the modified lattice and using the correspondance between the FRT to $T$ 
and the FPT from $S$, to get a relation between the $n$-th moment of
the FPT and the  lower-order moments of the FPT: 
\begin{equation}
\left<\mathbf{T}^n\right>_{\bf r} = \frac1{j({\bf r})}\sum_{m=1}^n 
\sum_{{\bf r}' \neq {\bf r}_T} (-1)^{m+1} {n \choose m} \pi_{\bf r}({\bf r}')
\left<\mathbf{T}^{n-m}\right>_{{\bf r}'} \nonumber
\end{equation} 
We denote by $\pi_{\bf r}$ the stationary distribution of the modified 
graph whose starting point is ${\bf r}$. The lowercase ${\bf r}$ refers to the 
starting point of the walk. 
This allows one, by recurrence, to get an
estimation of the $n$-th order moment, for large
enough domains, but in the 3D case only (in fact, 
$H({\bf r}|{\bf r}')$ has to be negligible when ${\bf r}'$ is far from ${\bf r}$). After some calculations which will be detailed in a further publication, it can be shown that
 \begin{equation}
\left\langle \mathbf{T}^n\right\rangle = n!N^n\left[ \left( H_0 - H({\bf r}_S|{\bf r}_T)
\right)\left(H_0-\bar{H}\right)^{n-1} +\mathcal{O}(N^{-\frac23}) \right],
\label{highord}
\end{equation}
where  $H_0=H({\bf r}_T|{\bf r}_T)$ and $ \bar{H} = 1/N \sum_{{\bf r}} H({\bf r}|{\bf r}')$. Note that $ \bar{H}$ is  independent 
of ${\bf r}'$ due to the symmetry property of $H$, and that  $\bar{H}$ scales as 
$N^{-\frac13}$, since $G({\bf r}) \sim 1/r$.  A good estimation of $\bar{H}$, to be used for practical applications, is its value for a spherical domain, computed in the continuous 
limit, $\bar{H} = (18/5) 
(3/(4\pi))^{2/3}N^{-1/3}$.
The estimations (\ref{highord}) are confirmed by numerical simulations (fig. \ref{moments}). 

It should be pointed out that the moments (\ref{highord}) are close but not 
equal (see fig. \ref{moments}) to the moments of an exponential distribution 
of the FPT. 
However, if the
particle starts randomly inside
the volume, the moments are the same as those of the exponential 
distribution, with a correction proportional to $N^{-2/3}$. This property
sheds a new light on the quasi-chemical approximation 
\cite{bere}, 
which assumes that, if a particle starts randomly in the volume, it has a 
constant exit probability at each time step, which leads to an exponential 
distribution. 

\begin{figure}[t]
\centering\includegraphics[width = .7\linewidth,clip]{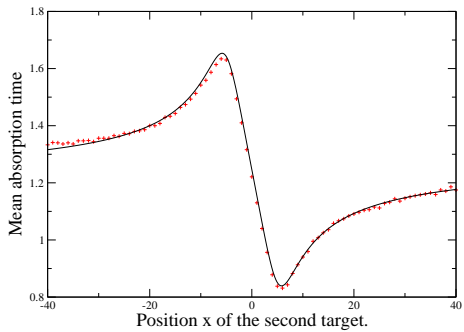}
\caption{2D: Two-target simulations.  Simulations (red crosses) 
vs. theory 
(plain line). One target is fixed at (-5,0); The source is fixed at (5,0); 
The other target is at (x,3). The domain is a square of
side 201, the middle is the point (0,0)}
\label{time2t}
\end{figure}

We now turn to the situation where the lattice contains several targets, relevant in many chemical applications \cite{Rice}. For sake of simplicity, the calculation is driven
 in the case of two targets, but may be easily extended to more absorbing points. We compute here the eventual hitting probability to a specified target $P_i$, the so-called "splitting probability" \cite{Redner}, as well as the mean time until the particle hits either of the two targets $\langle \mathbf{T}\rangle$.   
We modify the graph in the same way as in fig.\ref{astuce}: for both absorbing
points, denoted by  ${\bf r}_1$ and ${\bf r}_2$, the bonds relating 
them to their neighbours become one-way bonds, and a link is added from each 
target to the starting point ${\bf r}_S$. 
We denote $\pi({\bf r}_1) = j_1$, $\pi({\bf r}_2) = j_2$ and $ j = j_1 + j_2$.
Again, the relation  $\langle \mathbf{T}\rangle = 1/j-1$ provides the mean absorption
time, and the probabilities to hit ${\bf r}_1$ or ${\bf r}_2$,
are respectively $j_1/j$ and $j_2/j$. 
We obtain a relation analogous to (\ref{valpip}): 
\begin{equation}
\pi'({\bf r}) = \frac{1-j}N + j H({\bf r}|{\bf r}_S) - j_1 H({\bf r}|{\bf r}_1) - 
j_2 H({\bf r}|{\bf r}_2),
\end{equation}
then, writing $\pi'({\bf r}_1)=\pi'({\bf r}_2)=0$ 
\begin{equation}
\left\{ \begin{array}{rcl}
\frac{1-j}N + (j_1+j_2)H_{1s} - j_1H_{01} - j_2H_{12}
& = & 0 \\
\frac{1-j}N + (j_1+j_2)H_{2s} - j_2H_{02} - j_1H_{12}
& = & 0 \\
\end{array} \right.
\end{equation}
where $H_{12} = H({\bf r}_1|{\bf r}_2)$ and, for $i = 1$ or $2$,  
$H_{is} = H({\bf r}_i|{\bf r}_S)$, 
$H_{0i} = H({\bf r}_i|{\bf r}_i)$.
These equations yield exact expressions for the mean absorption time 
and the splitting probabilities, respectively: 
\begin{equation}
\left\{
\begin{array}{l}
\langle \mathbf{T}\rangle = N 
\frac{[H_{01}-H_{1s}][H_{02}-H_{2s}] - [H_{12}-H_{2s}][H_{12}-H_{1s}]}
{H_{01}+H_{02}-2H_{12}} \\
P_1 = \frac{H_{1s}+H_{02}-H_{2s}-H_{12}}{H_{01}+H_{02}-2H_{12}} \\
P_2 = \frac{H_{2s}+H_{01}-H_{1s}-H_{12}}{H_{01}+H_{02}-2H_{12}} \\
\end{array}
\right.
\nonumber
\end{equation}
Again, these expressions give excellent results when compared to simulations
(fig. \ref{time2t}).

We finally address the case of a continuous Brownian motion. The target $T$ is now a sphere  of 
radius $a$ 
centered on ${\bf r}_T$, and the Brownian particle has 
a diffusion coefficient $D$. It still starts from the point ${\bf r}_S$, at 
a distance $R$ from the center of the target. 
The results are quite similar to those obtained in the discrete
case, and the details of the computation will be published in a future paper.
The estimated MFPT within the infinite-space approximation are
\begin{equation}
\langle \mathbf{T}_{\mathrm{3D}}\rangle = \frac{V}{4 \pi D}\left( \frac1a - \frac1R \right) ;
\label{eqnt}
\langle \mathbf{T}_{\mathrm{2D}}\rangle = \frac{A}{2\pi D}\ln\frac{R}{a} 
\end{equation}
where $V$ and $A$ are the volume and area of the domains.
If the target is approximately centered, the uniform correction gives a
contribution to $\langle \mathbf{T}\rangle$ of $-R^2/(6D)$ in 3D and 
$-R^2/(4D)$ in 2D. 
The correction due to a flat reflecting boundary is the following: 
\begin{equation}
\left\{
\begin{array}{rcl}
\langle \mathbf{T}_{\mathrm{3D}}\rangle & = & \frac{V}{4 \pi D}
\left( \frac1a + \frac1{2d} - \frac1R -\frac1{R'} \right) \\
\langle \mathbf{T}_{\mathrm{2D}}\rangle & = & \frac{A}{2\pi D}\left(\ln\frac{R}{a}+
\ln\frac{R'}{2d}\right) \\
\end{array} \right.
\end{equation}
with $d$  the distance between the center of the sphere $T$ and the boundary, 
and $R'$  the distance between this same center and the reflexion of the 
starting point by the boundary. 
Note that these results significantly extend the 
(exact) formula of Pinsky \cite{Pinsky}, which only gives the leading term in 
$a$. 


In summary, we have presented here a new method of computation that yields very
accurate expressions of mean first-passage times for discrete random walks and continuous Brownian motion. These approximations have proven to be especially useful when the target is roughly in the middle of the bounded domain or near to a flat boundary.  
This approach also gives access to 
more complex quantities such as higher-order moments. These results may be of 
the greatest interest for systems involving diffusion in confined media.

\end{document}